\documentclass[12pt]{article}
\usepackage{graphics}
\begin{document}
\begin{center}
{The Coupling Constant of the Pseudovector Pion-nucleon Interaction \\ for the Magnetic Moment of Nucleon}
\end{center}
\begin{center}
{Susumu Kinpara}
\end{center}
\begin{center}
{\it Institute for Quantum Medical Science (QST) \\ Chiba 263-8555, Japan}
\end{center}
\begin{abstract}
The role of the suppression of the coupling constant in the pseudovector pion-nucleon interaction is examined
to account for the calculation of the magnetic moment of nucleon. 
Among three kinds of the higher-order corrections the vacuum polarization of the pion propagator contributes to the numerical value considerably.
On the other hand the lowest-order vertex correction for the pion-nucleon-nucleon vertex does not give large effect 
in the approximation of the on-shell momenta for nucleons.
\end{abstract}
\section*{\normalsize{1 \quad Introduction}}
\hspace*{4.mm}
Nucleon is the fundamental particle interacting with other particles by exchanging a lot of bosons.
Especially pion generates the strong force and the virtual processes around the nucleon modify the dynamical properties of the propagation.
Supposing that pion plays a significant role to describe the nuclear system the field theoretical method is necessary to examine the reaction mechanism.
Including the effects of the interaction the propagators are required to extract the constants of the structure inherent to the phenomena.
\\\hspace{4.mm}
One of the subjects which attracts our interest is the electromagnetic interaction and the magnetic moment of nucleon.
The cloud of pion surrounding the bare nucleon gives rise to an additional effect which is to be included by the perturbative calculation
to construct the anomalous part in conjunction with the self-energy 
since the strength of the nuclear force is stronger than the electromagnetic interaction related with photons.
\\\hspace{4.mm}
The virtual pions also interact with photon and the anomalous part consists of the pion and the nucleon current parts.  
Both of the results of the lowest-order corrections are found to be convergent about the integral related to the diagrams 
with the pseudovector coupling $\pi$-N interaction.
In addition to the result of the calculation which is convergent the numerical values explain the experimental data well
provided the coupling constant of the $\pi$-N interaction $f$ is reduced about 20$\%$ from the standard value \cite{Kinpara}.
The present study examines the source of the shift of the coupling constant on the basis of the next-order perturbative calculation.
\section*{\normalsize{2 \quad The vertex correction in the pseudovector $\pi$-N interaction}}
\hspace{4.mm}
The pion-nucleon-nucleon three-point vertex is corrected same as the case of the photon-nucleon-nucleon vertex by using the pion propagator.
In momentum space the Feynman rule makes us write the vertex explicitly as 
\begin{eqnarray}
\qquad\qquad \Gamma_0 (p^\prime, p) = - \frac{f}{m} \tau_a \gamma_5 \gamma \cdot q     \qquad\qquad (q \equiv p^\prime - p)
\end{eqnarray}
for the lowest-order in the pseudovector coupling interaction with the coupling constant $f$ (=1) and the mass of pion $m$. 
Through the calculation of the higher-order the incoming and the outgoing momenta $p$ and $p^\prime$ are restricted to the on-shell state and the vertex function
is put between the Dirac spinors so that it is dependent solely on the momentum transfer $q^2$.
\\
\hspace{4.mm}
Using the rule of the diagram in momentum space the vertex function of the second-order is given as
\begin{eqnarray}
\Gamma_2 (p^\prime, p) = (\frac{f}{m})^3 \int \frac{d^4k}{(2\pi)^4} \,i\Delta_0(k)\,\tau_j\gamma_5\gamma\cdot(-k)\,iG_0(p^\prime-k)\nonumber\\
\times \,\tau_a\gamma_5\gamma\cdot(-q)\,iG_0(p-k)\,\tau_j\gamma_5\gamma\cdot k
\end{eqnarray}
where the $\Delta_0$ and $G_0$ are the free pion and nucleon propagators. 
The non-perturbative term has not been added in the present study following the fact that the calculation of the anomalous interaction also does not use the term.
The integrand in Eq. (2) has the order $k \sim -2$ and it is not apparent that the renormalization is possible or not 
different from the pseudoscalar interaction which is $k \sim -4$ and divergent only logarithmically.
\\\hspace{4.mm}
The $k$-integral is performed by the dimensional reguralization method using the related formulas 
and which makes the original form convert to the integral on the parameters $s$ and $t$ as  
\begin{eqnarray} 
\Gamma_2 (p^\prime, p) = - \frac{f}{m} \tau_a \gamma_5 \gamma \cdot q \, \rho \, (a_0 + a_1 \frac{q^2}{M^2} + a_2 (\frac{q^2}{M^2})^2 +O((q^2)^3))
\end{eqnarray}
\begin{eqnarray}
\rho \equiv \frac{M^2}{(4\pi)^2}(\frac{f}{m})^2
\end{eqnarray}
\begin{eqnarray} 
a_i \equiv \frac{1}{2} \int_0^1 ds \int_{-s}^s dt \; c_i(s,t)
\end{eqnarray}
\begin{eqnarray} 
c_0(s,t) = 2(\eta-1-{\rm log}\frac{M^2}{m^2}-2 \,{\rm log} s)(1+6s^2)-4 + 4 s +6 s^2
\end{eqnarray}
\begin{eqnarray} 
c_1(s,t) = -\frac{1}{2} -s + t +\frac{3}{2}(s^2 -t^2) +\frac{t^2}{2 s^2} \qquad\qquad\qquad\qquad  \nonumber\\
 \qquad\qquad + (\eta-1-{\rm log}\frac{M^2}{m^2}-2 \,{\rm log} s)\{1-3(s^2-t^2)\}
\end{eqnarray}
\begin{eqnarray} 
c_2(s,t) = \frac{1}{4 s^4} \{ s^4 - t^4 +\frac{1}{4}(1-6 s^2)(s^2-t^2)^2 \}
\end{eqnarray}
\begin{eqnarray} 
\eta \equiv \frac{2}{\epsilon}-\gamma +1 -{\rm log}\frac{m^2}{4 \pi \mu^2}
\end{eqnarray}
up to the $O((q^2)^2)$ order in the series of $q^2 / M^2$ where $M$ is the nucleon mass.
The definitions of the quantities $\epsilon$, $\gamma$ and $\mu$ in Eq. (9) are seen in Ref. \cite{Kinpara}.
It is noted that the expansion in $q^2 / M^2$ is correct within the radius of the convergence.
The coefficient $a_1$ is convergent and we do not need to subtract the divergent term beyond the constant term. 
The numerical values are $a_1 =0$ and $a_2 =1/15$ in the approximation neglecting the mass of pion $m^2 \approx 0$.
Although the $a_0$ is divergent it is not essential to derive the vertex function since the renormalization enables to eliminate the term legitimately.  
\\\hspace{4.mm}
The dependence of $a_1$ on the pion mass $m$ is thought to be small as shown below.
The correction to $c_1(s,t)$ is denoted by $c_1^\prime(s,t)$ and which is given as
\begin{eqnarray} 
c_1^\prime(s,t) = \frac{\lambda(1-s)}{s^2+\lambda(1-s)} \cdot \frac{s^2-t^2}{4} \{\, \frac{2(1-3 s^2)}{s^2} -\frac{\lambda(1-s)}{s^2+\lambda(1-s)} \,\} \nonumber\\
\nonumber\\
- \{ 1-3 (s^2-t^2) \} \,{\rm log} [ 1+\lambda s^{-2} (1-s) ]  \qquad
\end{eqnarray}
While $\lambda \equiv m^2/M^2$ is small ($\lambda \sim 0.02$) the expansion in $\lambda$ is not possible at $s \sim 0$ and the $s,t$-integral is done numerically 
and it yields $a_1^\prime = -0.023$.
The $m^2 \neq 0$ effect on $a_2$ has not been calculated for the $q^2-m^2$ term arising from the $\sim (q^2)^2$ in Eq. (3) is suppressed by the factor $\lambda$.
\\\hspace{4.mm}
Taking into account the vertex correction the coupling constant $f$ is modified as $f \rightarrow f f_1(q^2)$ with
\begin{eqnarray} 
f_1(q^2) = 1 + \rho \, [\, (a_1+2 \lambda a_2) \frac{q^2-m^2}{M^2} + a_2 (\frac{q^2-m^2}{M^2})^2 + O((\frac{q^2-m^2}{M^2})^3)\,]
\end{eqnarray}
effectively.
The subtraction of the constant term is determined so as to satisfy the renormalization condition $f_1(m^2) = 1$ about the external line of pion.
At the region $q^2 \sim M^2$ typical for the use of the pion propagator the factor does not shift so much as $f_1(M^2) = 1.013$.
It does not necessarily mean that the the vertex corrections have an influence on the momentum dependence of the effective
coupling constant only weakly since the result is obtained under the on-shell condition for both nucleons.
In fact the argument $q^2$ generally exceeds the radius of the convergence and the series is not convergent.
\section*{\normalsize{3 \quad The higher-order effects to the pion and nucleon propagators  }}
\hspace{4.mm}
In the calculation of the anomalous part of the magnetic moment the internal line of the pion propagator is  
changed from the non-interacting one as $\Delta_0(q) \rightarrow \Delta(q)$.
It is expressed by the Dyson equation and the solution is given as $\Delta(q) = (\Delta(q)^{-1} - \Pi(q))^{-1}$ using the polarization function $\Pi(q)$.
So our interest is to determine the form of $\Pi(q)$ by the perturbative calculation of the vacuum polarization.
It has been verified that the non-perturbative term in the pseudovector coupling does not contribute to $\Pi(q)$.
In the present study the form of the lowest-order approximation $\Pi(q) \approx \Pi_0(q)$ is applied 
to give the correction.
\\\hspace*{4.mm}
The factor $f_3(q^2)$ of the modification $f \rightarrow f f_3(q^2)$ 
due to the vacuum polarization is given as
\begin{eqnarray} 
f_3(q^2) = (1-\frac{\Pi_0 (q)}{q^2-m^2})^{-\frac{1}{2}}
\end{eqnarray} 
applying the renormalized form of $\Pi_0 (q)$.
One of the examples of the effective coupling constant is the anomalous interaction in the electromagnetic process.
The diagram consists of two parts that is the part of the pion current and the nucleon current 
and their higher-order corrections are represented by the common factor $f_3(q^2)^2$
approximately when the momentum of the virtual photon approaches to zero.
\\\hspace*{4.mm}
The renormalized form of $\Pi_0(q)$ has been calculated and it is given as follows
\begin{eqnarray} 
\Pi_0(q) = \frac{1}{\pi^2}(\frac{f}{m})^2 M^2 \,[\, q^2 (I(q) - I_0) - m^2 \,I_1 (q^2-m^2) \,]
\end{eqnarray} 
\begin{eqnarray} 
I(q) \equiv \int_0^1 d z \,{\rm log} \,[1-\frac{z(1-z) q^2}{M^2}]
\end{eqnarray} 
where $I_0$ and $I_1$ are the coefficients of the expansion in the vicinity of $q^2=m^2$ such as $I(q) = I_0 + I_1 (q^2-m^2) + O((q^2-m^2)^2)$.
It corresponds to the polarization function in the medium of the nuclear matter at the limit of the zero density of nucleon 
with $m \rightarrow 0$ about the pion mass \cite{Serot}.
At $q^2 = M^2$ typical of the momentum for the pion propagator the $f_3(q^2)$ gives $f_3(M^2) =0.74$ suitable 
to examine the effect of the decrease of the coupling constant $f$.
\\\hspace*{4.mm}
An interesting property of the pseudovector coupling interaction is the non-perturbative term which is ascribed to the $\theta$-function in
the time-ordered product.
It is generated from the non-perturbative relation by using the equation of motion for the $\pi$-N system
besides the cancellation between the normal-dependent term and the contact interaction in the calculation of the perturbative expansion. 
Imposing the on-shell condition the non-perturbative term in the vertex vanishes and then the vertex correction and the vacuum polarization mentioned above
are applicable to the higher-order corrections by the renormalized functions.
\\\hspace*{4.mm}
The calculation of the self-energy for the nucleon propagator encounters the difficulty preparing the finite result under the usual manner
of the renormalization by the counter terms of the mass and the field because of the derivative coupling.
In such a case the non-perturbative term plays a decisive role to construct the self-energy and enables to obtain the convergent result.
The divergent parts occurring in the dimensional reguralization integral are cancelled in the fraction along with the usual procedure of the counter terms. 
\\\hspace*{4.mm}
Determining the self-energy $\Sigma(p)$ the renormalized propagator of nucleon $G(p)$ is expressed
\begin{eqnarray} 
G(p) = \frac{1}{\gamma\cdot p -M -\Sigma(p)} = \frac{\alpha(p^2)\gamma\cdot p + \beta(p^2) M}{p^2 -M^2}
\end{eqnarray} 
\begin{eqnarray} 
\Sigma(p) = M c_1(p^2) - \gamma \cdot p \, c_2(p^2)
\end{eqnarray} 
in two kinds of the expressions.
These four quantities $c_i(p^2)$ ($i$=1,2), $\alpha(p^2)$ and $\beta(p^2)$ are not independent and connected by Eq. (15) in conjunction with 
the renormalization conditions.
It is given by the relations between the coefficients of the expansion in the series of $p^2-M^2$.
For example in the case of $c_i^{(n)}$ such that $c_i(p^2) = \sum_{n=0}^{\infty} n!^{-1} c_i^{(n)} (p^2-M^2)^n $
the renormalization condition imposes the relations $c \equiv c_1^{(0)} = c_2^{(0)}$ and $c_1^{(1)}-c_2^{(1)} = c_2^{(0)} /2 M^2$
and $\alpha(M^2) = \beta(M^2) = 1$ is satisfied to construct the rationalized form of $G(p)$.
\\\hspace*{4.mm}
It is useful to express the coefficients of the expansion $\alpha^{(n)}$ and $\beta^{(n)}$ in terms of $c_i^{(n)}$ 
at the region around the on-shell point $p^2 = M^2$.
The coefficients of the rationalized form of $G(p)$ has been used to carry out the perturbative calculation 
along with the method of the inversion of the matrix for processes related to the external pions.
The relation is presented up to $n=2$ order as 
\begin{eqnarray} 
\alpha^{(1)} = \frac{c^2}{4 M^2 (1+c)} -c_2^{(1)} +M^2(c_1^{(2)}-c_2^{(2)}) \qquad\quad
\end{eqnarray} 
\begin{eqnarray} 
\beta^{(1)} = \alpha^{(1)} + \frac{c}{2 M^2 (1+c)}  \qquad\qquad\qquad\quad \qquad\quad
\end{eqnarray} 
\begin{eqnarray} 
&&\alpha^{(2)} = 2 \alpha^{(1) \,2} -\frac{c^2 \, c_2^{(1)}}{2M^2 (1+c)^2} +\frac{c\,(c_1^{(2)}-c_2^{(2)})}{1+c} -c_2^{(2)} \quad\qquad \nonumber\\
&&+\frac{2 M^2}{3} (c_1^{(3)}-c_2^{(3)})
\end{eqnarray}
\begin{eqnarray} 
\quad \beta^{(2)} = \alpha^{(2)} + M^{-2} \alpha^{(1)}  + \frac{1}{M^4 (1+c)^2}(-\frac{c^2}{4} + M^2 c_2^{(1)}) 
\end{eqnarray} 
In our previous study $\alpha^{(1)}$ and $\beta^{(1)}$ has been applied to calculate the cross section of the photoproduction of $\pi^{+}$ 
from the threshold to the resonance energy region.
The approximate value of the self-energy is determined so as to reproduce the low-energy pion-nucleon elastic scattering.
Meanwhile the $n=2$ model is appropriate 
since the magnetic moment of nucleon is calculated by using the values ($c = 2$, $\,c_1^{(1)}=M^{-2}$, $c_2^{(1)} = c_i^{(n \ge 2)} =0$). 
It yields $\beta^{(1)} = 2 \alpha^{(1)} = 2 / 3 M^2 $ and $\beta^{(2)} = 2 \alpha^{(2)} = 4 / 9 M^4 $.
The expansion of $\alpha(p^2)$ and $\beta(p^2)$ at the on-shell point does not have difficulty taking into account the complex poles.
The prescription follows the way of the perturbative calculation in which the poles out of the real axis are neglected \cite{Feldman}. 
\\\hspace*{4.mm}
Different from above two corrections the nucleon propagator does not supply the factor multiplied by the free propagator.
When the approximation $\alpha(p^2) \approx \beta(p^2)$ is suitable also for the off-shell regions 
it is possible to create the multiplicative factor $f_2(p^2)$ such as $\sqrt{\alpha(p^2)}, \,\sqrt{\beta(p^2)}$ and $\sqrt{(\alpha(p^2)+\beta(p^2))/2}$
according to the definition of it.
While three possible factors are separated from each other except the on-shell point they all increase monotonously as $p^2$ goes to infinity. 
It indicates that the factor plays a role to restore too strong effect of the vacuum polarization in the region of the nucleon momentum $p^2 < M^2$.
\\\hspace*{4.mm}
The value of the pion momentum $q^2 = (p^\prime -p)^2 \sim M^2$ has been introduced tentatively.
The choice is not only an assumption and expected by the consideration of the $k$-integral
to obtain the self-energy of the nucleon in the lowest-order calculation.
The denominator of the integrand has the zero 
at the specific values $\bar{k}^\prime= \pm \, p z$ of the variable $k^\prime$ transformed from the momentum $k$ of the internal nucleon by $k^\prime \equiv k + p z$
where $p$ is the momentum of the on-shell nucleon. 
The parameter $z$ of the Feynman formula of the integral is replaced with the average $\bar{z} =1/2$.
According to the sign of $\bar{k}^\prime$ the $\bar{k}^2$ has the values 0 or $M^2$ 
and the former is chosen as $\bar{k}^2$ for nucleon.
\\\hspace*{4.mm}
The same procedure is applied to $k^2$ of the pion propagator for the internal line by changing as $M \leftrightarrow m$ and $p \rightarrow -p$
and it results in $M^2$ or 0 according to $\bar{k}^\prime= \pm \, p z$.
To derive these relations the term of the pion mass $m^2$ is dropped approximately.
The numerical values of these factors are $f_3(M^2)=0.74$ and $f_2(0)=0.88$ for $\sqrt{\alpha(0)}$.
The use of the upper sign for $\bar{k}^\prime$ ($\bar{k}^\prime= + \, p z$) is consistent 
with the integral of the nucleon current part in the calculation of the anomalous magnetic moment as described below.
\\\hspace*{4.mm}
It is important to search for the factor $f_3$ in another way to understand the suppression of the effective coupling constant.
Regarding to the nucleon current part the shift of the variable $k \rightarrow k^\prime = k - \alpha $ is same as that of the integral in Eq. (2).
The variable $k^\prime$ may be represented by the parameter $\alpha \equiv p x + p^\prime y$ with the average values $x = y =1/3$
and $\bar{k}^\prime = +\alpha$ is chosen since the integrand is transformed to $ \sim ({k^\prime}^2 -\alpha^2)^{-1}$.
Using it $\bar{k}^2 \sim 4 \alpha^2$ is deduced.
Substituting $q^2 = k^2$ the average value of $\alpha^2$ is $\alpha^2 \sim 4 M^2 /9 -{\bar k}^2 /9$.
The joint use of these relations yields the expected value $\bar{k}^2$ of the pion momentum $k$ as $\bar{k}^2 \sim 1.23 M^2$
and the factor results in $f_3(\bar{k}^2)=0.69$.
\\\hspace*{4.mm}
The numerical value of the factor is expected to be close to that extracted from the experimental value.
The coefficients of the anomalous magnetic moments of proton ($\kappa_p$) and neutron ($\kappa_n$) are connected with the self-energy stemmed from
the non-perturbative term and the perturbative terms of the vertex corrections 
by virtue of the Ward-Takahashi identity for the case of the photon-nucleon-nucleon vertex as 
\begin{eqnarray} 
\kappa_p \tau_p + \kappa_n \tau_n = z \tau_p + a \tau_3 + b ( 1 + \tau_n ) 
\end{eqnarray} 
in terms of the third component of the isospin matrix $\tau_3$ and $\tau_{p/n} \equiv (1 \pm \tau_3)/2$.
Then the vertex corrections of the part of the pion current ($a$) and the nucleon current ($b$) are given as 
\begin{eqnarray} 
a = ( -2 z + 2 \kappa_p - \kappa_n ) / 3
\end{eqnarray} 
\begin{eqnarray} 
b = ( - z + \kappa_p + \kappa_n ) / 3
\end{eqnarray} 
Using the parameter $z = 2(1+\lambda)^{-1}$ by the $n=2$ model (the lowest-order in $\gamma\cdot p -M$) for the self-energy 
and the experimental values of the anomalous magnetic moment
the $a$ and $b$ are required to be $a = 0.528$ and $b = -0.692$.
On the other hand the perturbative calculations of the vertex corrections become $a^\prime = 0.831$ and $b^\prime = -1.086$ 
and the expected values of the factor $f^\prime$ ($f \rightarrow f f^\prime$) result in 0.797 and 0.798 to reproduce $a$ and $b$ respectively.
\\
\section*{\normalsize{4 \quad Summary and remarks }}
\hspace*{4.mm}
Three corrections have been investigated to explain the role of the suppression of the coupling constant 
for the anomalous part of the magnetic moment. 
The perturbative lowest-order calculation of the vertex correction is convergent after the renormalization of the coupling constant
however the amount of the vertex function is not enough to observe the considerable shift of the factor.
The exact treatment is necessary to obtain the true value since it is not realistic to think 
that the result is unchanged when the on-shell condition is not acceptable for the vertex connected to the internal fermion line.
As for the pion propagator the polarization function is found to possess an interesting property at the large momentum transfer regions.
Because of the radius of the convergence the expansion is not available to examine the contribution of the lowest-order alone.
The factor is necessarily defined by the series in the inverse of the coupling constant when the degree of the off-shell is large.
\\
\hspace{4.mm}

\end{document}